\documentclass[amssymb,amsmath,aps,showpacs,floatfix,nofootinbib,showpacs,12pt]{revtex4}
\def\beq{\begin{equation}}
\def\eeq{\end{equation}}
\def\be{\begin{equation}}
\def\ee{\end{equation}}
\def\bea{\begin{eqnarray}}
\def\eea{\end{eqnarray}}
\def\nnb{\nonumber}

\newcommand{\gsim}{\lower.7ex\hbox{$\;\stackrel{\textstyle>}{\sim}\;$}}
\newcommand{\lsim}{\lower.7ex\hbox{$\;\stackrel{\textstyle<}{\sim}\;$}}

\begin{document}

%\begin{center}
 %\vspace{0.2cm}
 \title{ keV scale $\nu_R$ dark matter and its detection in $\beta$ decay experiment}
 \author{ Wei Liao}
 \affiliation{
  Institute of Modern Physics,
 East China University of Science and Technology, \\
 130 Meilong Road, Shanghai 200237, P.R. China %\\
 \vskip 0.1cm
 Center for High Energy Physics, Peking University,
 Beijing 100871, P. R. China
}

%\end{center}

\begin{abstract}
 %\vskip 0.2cm
 We study dark matter(DM) in the model with one keV scale right-handed
 neutrino $\nu_{R1}$ and two GeV scale right-handed neutrinos $\nu_{R2,3}$,
 the $\nu$SM. We find that one of the GeV scale right-handed neutrinos can
 have much longer lifetime than the other when two GeV scale right-handed
 neutrinos are degenerate. We show that mass and mixing of light neutrinos
 can be explained in this case. Significant entropy release can be generated
 in a reheating produced by the decay of one of the GeV scale $\nu_{R}$.
 The density of $\nu_{R1}$ DM can be diluted by two orders of magnitude
 and the mixing of $\nu_{R1}$ with active neutrinos is allowed to be
 much larger, reaching the bound from X-ray observation. This mixing can
 lead to sizeable rate of $\nu_{R1}$ capture by radioactive nuclei. The
 $\nu_{R1}$ capture events are mono-energetic electrons with keV scale energy
 away from the beta decay spectrum. This is a new way to detect DM in the
 universe.
\end{abstract}
 \maketitle

 {\bf Introduction:}
 Among many models of DM candidate a seesaw type model, the $\nu$SM, is
 of particular interests~\cite{ABS}. In this model there is one keV scale right-handed
 neutrino $\nu_{R1}$ and two GeV scale right-handed neutrinos $\nu_{R2,3}$. The keV scale
 $\nu_{R1}$ has a lifetime longer than the age of the universe and is a DM candidate.
 One of the virtue of this model is that the DM particle is
 already introduced in seesaw mechanism for explaining the tiny masses
 of active neutrinos.

 One interesting implication of $\nu_{R1}$ DM is that $\nu_{R1}$
 can interact with Standard Model(SM) particles through small mixing with active neutrinos
 and it might be detected in laboratory experiments. However it was
 shown by some authors ~\cite{bhl} that astrophysical constraints and the constraints
 from the mass and mixing of light active neutrinos are very strong.  Mixing of $\nu_{R1}$
 with active neutrinos is very small and detecting $\nu_{R1}$ in
 laboratory experiment seems difficult.
 In this article we investigate the mixing of $\nu_{R1}$ and its density
 in the thermal history of the early universe. We show
 that relatively large mixing of $\nu_{R1}$ with active neutrinos
 is allowed in the $\nu$SM when two GeV scale $\nu_{R2,3}$ are
 degenerate. We study the direct detection of
 the cosmic background of this keV scale $\nu_{R1}$ DM in
 beta decay experiment. In the following we briefly describe the $\nu$SM and discuss
 issues related to DM. We study the mixing of light neutrinos with heavy neutrinos
 in the case that two GeV scale right-handed neutrinos are degenerate. We analyze the capture
 of the cosmic background of $\nu_{R1}$ DM
 in target of radioactive nuclei $^3$H and $^{106}$Ru.

 {\bf keV scale $\nu_{R1}$ DM in $\nu$SM:}
 In seesaw mechanism mass matrix of active neutrinos is given by
 \bea
 m_{\nu} =v^2 Y^* M_R^{-1} Y^\dagger, \label{see-saw2}
 \eea
 where $M_R$ is the Majorana mass matrix of right-handed
 neutrinos, $Y$ the matrix of Yukawa couplings, $v$ the vacuum expectation
 value of doublet Higgs in the SM: $<H>=(0,v)^T$.
 In seesaw mechanism $Y$ can be parameterized as~\cite{ci}
 \bea
 Yv= U ({\tilde m}_\nu)^{1/2} P^{1/2} O (M_R^*)^{1/2}, \label{see-saw3}
 \eea
 where ${\tilde m}^{1/2}_\nu =\textrm{diag} \{ m_1^{1/2} , m_2^{1/2}, m_3^{1/2} \}$
 and $P^{1/2} =\textrm{diag} \{e^{i\phi_1/2},e^{i\phi_2/2},e^{i\phi_3/2}
 \}$.
 Real numbers $m_i(i=1,2,3)$ are the masses of light neutrinos and
 $\phi_i(i=1,2,3)$ are Majorana phases.
 $(M^*_R)^{1/2} =\textrm{diag}\{(M^*_1)^{1/2}, (M^*_2)^{1/2}, (M^*_3)^{1/2}\}$.
 $U$ is the neutrino mixing matrix observed in neutrino oscillation experiments. $O$ is
 a complex orthogonal matrix: $O^T O=O O^T=1$.

 Active neutrinos mix with heavy Majorana neutrinos through a mixing matrix
 \bea
 R=Y v (M^*_R)^{-1}. \label{mixing2}
 \eea
 Using Eq. (\ref{see-saw3}) $R$ is rewritten as
 \bea
 R= U ({\tilde m}_\nu)^{1/2} P^{1/2} O (M_R^*)^{-1/2}. \label{mixing3}
 \eea
 For convenience we will work in the base that $M_{1,2,3}$ are real.

 keV scale $\nu_{R1}$ can decay to light active neutrinos or
 photon and its main decay channel is the decay to three active
 neutrinos. For $\theta^2_1=\sum_l |R_{l1}|^2 \sim 10^{-8}$ the
 lifetime of $\nu_{R1}$ is estimated $\sim 10^{21}$ s ~\cite{ABS}
 which is much larger than the age of the universe $ \sim 10^{17}$s
 and long enough for a DM candidate.
 $\nu_{R1}$ DM can be produced in the early universe through
 active-sterile neutrino oscillation~\cite{DW,ABS,others,blrv}.
 For $\theta_1^2 \sim 10^{-8}$ significant amount of
 keV scale $\nu_{R1}$ can be produced. With larger mixing
 $\nu_{R1}$ DM can be over-produced using this mechanism.
 Many other aspects of this model of DM, e.g. constraints from X-ray observation
 and Lyman-$\alpha$ forest, detection of this DM, possible symmetries
 etc., have been analyzed~\cite{bnrst,bnr,birs,blrv2,bri,BS,ak,sh,hll}.
 Suggestions beyond the minimal model have also been
 made and other DM production mechanism has been considered~\cite{ST,Kusen,PK,KS,Gouvea}.

 $\nu_{R1}$ density can be diluted in the early universe and
 $\theta_1^2$ is allowed to be much larger if significant entropy release
 is produced in a reheating at MeV temperature scale before the Big Bang
 Nucleosynthesis(BBN). This can be achieved by
 the decay of a non-relativistic particle in the early universe.
 It's interesting that GeV scale right-handed neutrino is a candidate
 of this non-relativistic particle. However it was noticed that
 this is a further constraint on the seesaw mechanism and large enough
 entropy production can not be obtained because of the constraint
 from mass and mixing of light neutrinos ~\cite{bhl}.
 In the following we show that
 the analysis in ~\cite{bhl} does not really apply to the case when two GeV scale
 right-handed neutrinos are degenerate ($M_3=M_2$) and
 $\theta_1^2 \gg 10^{-8}$ is indeed allowed.

{\bf Entropy production by GeV scale $\nu_R$ and neutrino mixing:}
 Using an explicit example we show that mass and mixing of light
 neutrinos and enough entropy production by the decay of a GeV right-handed
 neutrino can all be accommodated in $\nu$SM when two GeV scale right-handed
 neutrinos are degenerate.

 Two degenerate right-handed neutrino states
 can be rewritten in other base using a unitary transformation:
 \begin{eqnarray}
 \begin{pmatrix} \nu'_{R1} \cr \nu'_{R2} \cr \nu'_{R3} \end{pmatrix} =
 \begin{pmatrix} 1 & 0 & 0 \cr 0 & c & -s \cr 0 & s^* & c^* \end{pmatrix}
 \begin{pmatrix} \nu_{R1} \cr \nu_{R2} \cr \nu_{R3} \end{pmatrix},
 \label{transform}
 \end{eqnarray}
 where $|c|^2+|s|^2=1$. The symmetric mass matrix $M'_R$ in this base is no longer diagonal.
 We note that two states $|\nu'_{R2,3}>$ have the same mass of
 $|\nu_{R2,3}>$. This is just another way to write the mass term. For
 example, if $c=1/\sqrt{2}$ and $s=i/\sqrt{2}$, $M'_R$ is
 obtained (up to a phase factor $i$ ) as
 \begin{eqnarray}
  M'_R = \begin{pmatrix} M_1 & 0 & 0 \cr
                          0 & 0 & M_2 \cr
                          0 & M_2 & 0 \cr
 \end{pmatrix}.
 \label{massterm}
 \end{eqnarray}
 In Eq. (\ref{massterm}) $M_2$ is written in the form of Dirac type.
 Another way to see this point is to note that $M_R M_R^\dagger$
 and $M_R^\dagger M_R$ are not changed by Eq. (\ref{transform})
 when $M_3=M_2$. Hence Eq. (\ref{transform}) does no change the energy dispersion
 although $M_R$ is transformed by it.

 When $\nu_{R2,3}$ are degenerate we should transform them to
 interaction base to understand their interaction with active neutrinos.
 In the interaction base of $\nu'_{R2,3}$ $Y'$ is obtained from
 $Y$: $Y'=Y V^\dagger$ where $V$ is the unitary matrix
 in $\nu'_R=V \nu_R$ as shown in Eq. (\ref{transform}).
 In the interaction base $Y^{'\dagger} Y'$ is obtained from $Y^\dagger Y$ to
 the following form
 \bea
 Y^{'\dagger} Y'= \begin{pmatrix} y_1^2 & y_{12}^2 & y_{13}^2 \cr
               y_{12}^{2*} & y_2^2 & 0 \cr
               y_{13}^{2*} & 0 & y_3^2
 \end{pmatrix}.
 \label{Yprime}
 \eea
 Coupling of $\nu'_{R2,3}$ with active
 neutrinos are $y_{2,3}$ times active neutrino mixing.

 As an example, we consider normal mass hierarchy and matrix $O$ of
 the following form
 \begin{eqnarray}
 O = \begin{pmatrix} 1 & 0 & 0 \cr
                          0 & c_\theta & s_\theta \cr
                          0 & -s_\theta & c_\theta \cr
 \end{pmatrix},
 \label{NHmatrix}
 \end{eqnarray}
 where $c_\theta=\cos\theta$, $s_\theta=\sin\theta$ and $\theta=x+iy$.
 $x$ and $y$ are real numbers.
 Diagonalizing the second and third entries of $Y^\dagger Y$ using Eq. (\ref{transform})
 we obtain $c=\cos\alpha $, $s=\sin\alpha ~e^{i \beta} $ where
 \bea
 &&\tan2\alpha=-\frac{2D} {A-B}, \\
 && D e^{i\beta}=(m_2 c_\theta^* s_\theta-m_3 s_\theta^* c_\theta)M_2^{1/2}
 M_3^{1/2},
 \eea
 and $D=|(m_2 c_\theta^* s_\theta-m_3 s_\theta^* c_\theta)M_2^{1/2}
 M_3^{1/2}|$, $ A=M_2 (m_2 |c_\theta|^2+m_3 |s_\theta|^2)$, $B=M_3(m_2
 |s_\theta|^2+m_3 |c_\theta|^2)$.
  Three eigenvalues are found: $y_1^2=m_1 M_1/v^2$,
  \bea
  && y_2^2=\frac{2 C/v^2}{A+B+\sqrt{(A+B)^2-4 C}}, \\
  && y_3^2=\frac{1}{2 v^2}(A+B+\sqrt{(A+B)^2-4 C})
  \eea
  where $C=m_2 m_3 M_2 M_3$.

 For $|y| \gg 1$ it's easy to see that $|c_\theta|\approx |s_\theta| \approx
 \frac{1}{2} e^{|y|}$, $A \approx \frac{1}{4} M_2(m_2+m_3)
 e^{2|y|}$, $B \approx  \frac{1}{4} M_3(m_2+m_3)
 e^{2|y|}$ and we find
 \bea
 && y_2^2\approx \frac{4 C/v^2}{(m_2+m_3)(M_2+M_3)}e^{-2|y|}, \\
 && y_3^2 \approx \frac{1}{4 v^2}(m_2+m_3)(M_2+M_3)e^{2|y|}
 \eea
 For normal mass hierarchy we have $m_3\approx \sqrt{\Delta m^2_{atm}}\approx 0.05$ eV,
 $m_2\approx \sqrt{\Delta m^2_{solar}} \approx 0.009$ eV and $m_1 \ll m_2$.
 Using $M_3=M_2$ we get
 \bea
 y_2^2 \approx {2 m_2 M_2 \over v^2}  e^{-2|y|}, ~y_3^3 \approx {m_3 M_2 \over 2
 v^2} e^{2|y|}. \label{Yuka2}
 \eea
 In Eq. (\ref{Yuka2}) we see that coupling of $\nu'_{R2}$
 with active neutrinos are suppressed by $e^{-|y|}$ if $|y|$ is
 large and coupling of $\nu'_{R3}$ is enhanced. We note that
 $\cos 2\alpha \to 0$, $|\beta| \to \pi/2$ in the limit $|y| \to \infty$
 and the mass matrix $M'_R$ approaches the mass matrix shown in
 Eq. (\ref{massterm}).

 The mixing $R'$ in the interaction base can be similarly obtained.
 We obtain $\theta^{'2}_{1,2,3}=\sum_l |R^{'2}_{l1,l2,l3}|^2$ as
 \bea
 && \theta^{'2}_1=\frac{m_1}{M_1}, \\
 && \theta^{'2}_2 \approx \frac{2 m_2}{M_2} e^{-2 |y|}, \\
 && \theta^{'2}_3 \approx \frac{m_3}{2 M_2} e^{2 |y|}.
 \eea

 The decay rate of $\nu'_{R2}$ (for $M_2 < M_W$) is~\cite{bhl}
 \bea
 \Gamma=\frac{G^2_F M_2^5}{192 \pi^3} F \theta^{'2}_2 \approx\frac{G^2_F m_2 M_2^4}{96
 \pi^3} F e^{-2 |y|},~~
 \eea
 where a factor $2$ has been included to account for the charge conjugation processes
 and a factor $1/2$ has been included to account for the Majorana nature of  $\nu'_{R2}$.
 $F$ is a factor which accounts for effects of various final states~\cite{bhl}.
 $F\approx 16.0$ for $m_b \ll M_2 < M_W$. The lifetime of $\nu'_{R2}$ is
 \bea
 \tau\approx 0.44 ~\textrm{s} \times \frac{e^{2|y|}}{(2000)^2} ~\frac{0.01 ~\textrm{eV}}{m_2}
 ~\bigg( \frac{30 ~\textrm{GeV}}{M_2} \bigg)^4.~~
 \label{lifetime}
 \eea
 It is worth pointing out that $\nu'_{R2}$ can not decay through oscillation to $\nu'_{R3}$
 because the probability of oscillation to $\nu'_{R3}$ vanishes when $\nu'_{R2,3}$
 are degenerate. The above analysis and the later discussions
 can also be applied to quasi-degenerate $\nu_{R2,3}$ if their
 mass difference is so small that no significant oscillation can happen before $\nu'_{R2}$
 decays. For our whole discussion to be valid for quasi-degenerate $\nu_{R2,3}$,
 it's sufficient to assume that no significant oscillation can happen before the BBN time
 which is $\sim 1$ s. This condition is
 $|M_2-M_3| \times 1 ~\textrm{s} \ll \hbar$, or  $|M_2-M_3| << 6.58 \times 10^{-22}$ MeV.

 Assuming that $\nu'_{R2}$ is in thermal equilibrium at very high
 temperature and decouples when it is relativistic,
 the entropy release produced by the decay of $\nu'_{R2}$ is~\cite{stkt}
 \bea
 S && \approx 0.76 \frac{g_N}{2} \frac{{\bar g}_*^{1/4} M_2}{g_*
 \sqrt{\Gamma M_{Pl}}} \nnb \\
% && \approx 7. \times 10^{-3} \frac{ g_N {\bar
% g}_*^{1/4} e^{|y|} }{g_* \sqrt{G_F^2 m_2 M_2^2 M_{Pl}}}
 && \approx 96 \times \bigg( \frac{0.01 ~\textrm{eV}}{m_2} \bigg)^{1/2}
 \bigg( \frac{30 ~\textrm{GeV}}{M_2} \bigg) \times \frac{e^{|y|}}{2000},~~
 \label{entropy}
 \eea
 where $S=s_f/s_i$. $s_f$ and $s_i$ are entropies just after and
 before the decay of $\nu'_{R2}$.  $g_N=2$ is the number of degrees
 of freedom of $\nu'_{R2}$. $g_*$ is the effective number of degrees
 of freedom at $\nu'_{R2}$ freeze-out which is assumed at electro-weak scale.
 $g_* \approx 101.5$ when excluding top quark.
 ${\bar g}_*\approx 10.75$ is the effective number of degrees just
 after the entropy release.
 % $M_{Pl}=1.22 \times 10^{19}$ GeV.

 In Eq. (\ref{entropy}) we can see that $S \sim 100$ can be achieved for $e^{|y|}\sim 2000$.
 According to Eq. (\ref{lifetime}), for this range of parameter space $\nu'_{R2}$ has lifetime
 $\lsim  1$ s.  So it decays before the BBN and does not spoil
 the prediction of the BBN~\cite{ask}. It's also easy to see that
 without $e^{|y|}$ factor $M_2 \ll 1 $ GeV is needed to make $S$ larger
 than 100. However, $M_2 \ll 1 $ GeV is not allowed because
 it would give a too long lifetime to $\nu'_{R2}$ and it would spoil the prediction of
 the BBN. We emphasize
 that previous negative conclusion on the entropy
 production by $\nu_{R2}$~\cite{bhl} does not apply to the case we are
 considering. They worked in the base that mass matrix is
 diagonalized and did not pay attention to the fact that we should
 analyze the interaction of GeV scale right-handed neutrinos
 in their interaction base when they are degenerate.

 We point out that in our example $\theta^{'2}_1\sim 10^{-6}$ can be
 obtained by taking $m_1 \sim 10^{-3}$ eV which is allowed by the
 neutrino oscillation experiments. Astrophysical observation of
 the decay $\nu_{R1} \to \nu \gamma$ is able to constrain
 $\theta^{'2}_1$~\cite{brs}
 \bea
 \theta^{'2}_1 \lsim 1.8\times 10^{-5} \bigg(
 \frac{1~\textrm{keV}}{M_1} \bigg)^5. \label{X-ray}
 \eea
 We see that $M_1 \lsim 2$ keV is needed for $\theta^{'2}_1$ reaching $ \sim 10^{-6}$.
 $M_1$ is constrained by the observation of dwarf
 spherical galaxies: $M_1 \gsim 1-2$ keV~\cite{bri}. $M_1$ is
 also constrained by Lyman-$\alpha$ forest. In the
 case we are considering the Lyman-$\alpha$ bound is $M_1 \gsim 1.6$ keV when
 $S \approx 100$~\cite{bhl} which is re-scaled by a factor $S^{-1/3}$
 compared to the bound for non-resonant production of $\nu_{R1}$ DM~\cite{blrv2}.
 We further note that X-ray and Lyman-$\alpha$~\cite{blrv2} constraints become weaker when
  $\nu_{R1}$ DM accounts for part of the DM in the universe, e.g. $\sim 40\%$ of
  total DM energy density. In summary $M_1 \approx 2$ keV is allowed
  by the present constraints and $\theta^{'2}_1$ is allowed to reach $ \sim 10^{-6}$~\cite{bhl}.

 We note that $\nu'_{R2}$ has very small Yukawa coupling and
 can not come into thermal equilibrium due to this interaction.
 Other interaction of $\nu'_{R2}$ is needed to make it populated
 at temperature of electro-weak scale.
 Detailed model for it will be explored in future works.
 The example shown is a particular case with normal hierarchy.
 The quasi-degenerate mass pattern of light neutrinos would give
 too large mixing of keV scale right-handed neutrino with active neutrinos
 and is not compatible with the $\nu$SM~\cite{ABS,hll}.
 For inverted mass hierarchy, we can choose matrix $O$
 as
 \begin{eqnarray}
 O = \begin{pmatrix} 0 & \cos\theta & \sin\theta \cr
                          0 & -\sin \theta & \cos\theta \cr
                          1 & 0 & 0
 \end{pmatrix}.
 \label{IHmatrix}
 \end{eqnarray}
 Similar results can be obtained in the case that two GeV scale $\nu_{R2,3}$ are degenerate.

 {\bf $\nu_{R1}$ capture by radioactive nuclei:}
 $\nu_{R1}$ can interact with SM particles
 through mixing with active neutrinos. The four-Fermion
 interaction of $\nu_{R1}$ with electron is
 \bea
 \Delta L &&=\frac{G_F}{\sqrt{2}}R^*_{e1}{\bar n}\gamma^\mu(g_A-\gamma_5 g_A)
 p \nnb \\
 && \times {\bar \nu}_{R1}\gamma_\mu(1-\gamma_5) e + h.c., \label{Inter}
 \eea
 where $n$ and $p$ stand for neutron and proton. $g_{V,A}$ are vector
 and axial-vector form factors. According to Eq. (\ref{Inter})
 $\nu_{R1}$ DM in the universe can be captured by radioactive
 nuclei in processes
 \bea
 \nu_{R1}+A \to B^- +e^+~~\textrm{or}~~
 \nu_{R1}+A  \to B^+ + e^-. ~~~\label{proc}
 \eea
 The energy of electron produced in this process is
 \bea
 E_e= m_e+Q_\beta+M_1,
 \eea
 where $Q_\beta$ is the end point kinetic energy of the beta decay
 ($A \to B^+ + e^-+ {\bar \nu}_e$) or anti-beta decay ($A \to B^-+e^+ +\nu_e$).
 For $M_1 \ll Q_\beta$, the cross section of this process is
 similar to that of the capture of the cosmic relic neutrinos~\cite{cmm,lvv}.
 The capture of $\nu_{R1}$ DM produces mono-energetic electrons
 beyond the end point of the spectrum of beta decay or anti-beta decay.
 For super-allowed transition the cross section is
 \bea
 \sigma_{capt} v_\nu =\frac{\pi^2 ln2}{f t_{1/2}} ~p_e E_e F(Z,E_e) |R_{e1}|^2,
 \label{Xsection}
 \eea
 where $ p_e=\sqrt{2 m_e(Q_\beta+M_1)}$ and
 $F(Z,E_e)$ is the Fermi function. In Eq. (\ref{Xsection})
 a factor of $1/2$ arising from the Majorana nature of $\nu_{R1}$
 has been included. The cross section is
 normalized to the beta decay rate using $t_{1/2}$, the half-life
 of $A$ nucleus and the factor $f$
 \bea
 f=\int^{m_e+Q_\beta}_{m_e} F(Z,E'_e) p'_e E'_e E'_\nu p'_\nu dE'_e.
 \eea
 $p'_e$, $E'_e$ are momentum and energy of electron (or positron ).
 $p'_\nu$and $E'_\nu$ are the momentum and energy
 of anti-neutrino (or neutrino) in beta decay (or anti-beta decay).

 Consider $\nu_{R1}$ capture by Tritium: $\nu_{R1}+~^3H \to ~^3He+e^-$ which
 has $Q_\beta=18.59$ keV and $t_{1/2}=12.3$ year.
 Using parameters in \cite{cmm} we find that for
 $M_1 \ll Q_\beta$ the cross section of this process
 is $\sigma_{capt} v_\nu \approx 3.9
 \times 10^{-45}$ cm$^{-2} \times c \times |R_{e1}|^2 $
 where $c$ is the speed of light. The event rate,
 $\sigma_{capt} v_\nu n_{\nu_{R1}}$, is estimated
 \bea
 N \approx 0.71 ~\textrm{year}^{-1}\times \frac{n_{\nu_{R1}}}{10^{5} ~\textrm{cm}^{-3}}
 \frac{|R_{e1}|^2}{10^{-6}} \frac{^3 \textrm{H}}{10 ~\textrm{kg}}, \label{eventR}
 \eea
 where $\nu_{R1}$ DM is assumed to account for a significant part of
 DM and its number density is estimated
 \bea
 n_{\nu_{R1}} \sim
 10^5 ~\textrm{cm}^{-3}
 \frac{\rho_{\nu_{R1}}}{0.3 ~\textrm{GeV} \textrm{cm}^{-3}}
 \frac{3 ~\textrm{keV}}{M_1}. \label{numden}
 \eea
 In Eq. (\ref{numden}) we have used reference density $0.3$ GeV cm$^{-3}$,
 the estimated DM density in the galactic halo at the position of the solar system.
 This estimation of local density is extrapolated from the astrophysical observation
 of Milky Way and the model of DM halo which does not depend on whether the DM
 is warm or is cold. This estimation is applicable to warm DM.

 We find that $^{106}$Ru, which has $Q_\beta=39.4$ keV and $t_{1/2}=373.6$ days,
 is also good to detect $\nu_{R1}$.
 The cross section of $\nu_{R1}$ capture by $^{106}$Ru is
 $\sigma_{capt} v_\nu \approx 2.94 \times 10^{-45}$ cm$^{-2}
 \times c \times |R_{e1}|^2 $ for $M_1\ll Q_\beta$.
 The event rate of $\nu_{R1}$ capture is
 \bea
 N \approx 16 ~\textrm{year}^{-1} \times \frac{n_{\nu_{R1}}}{10^{5} ~\textrm{cm}^{-3}}
 \frac{|R_{e1}|^2}{10^{-6}} \frac{^{106} \textrm{Ru}}{10
 ~\textrm{Ton}}.~~ \label{eventR1}
 \eea
 We note that the production rate of $^3$H in reactor is $0.01\%$
 and the production rate of $^{106}$Ru is $0.4\%$. If $10$ kg Tritium are
 produced per year in reactors, around $12$ Tons of $^{106}$Ru are
 produced. We note that the reference number Eq. (\ref{numden}) used in Eq. (\ref{eventR1}) is a
 conservative estimate of the local number density. It's possible that
 the solar system is located in a sub-halo in which the local
 DM density is several orders of magnitude larger than the galactic value.
 If this happens, capture rate is several orders of magnitude
 larger.

 We emphasize that the events of $\nu_{R1}$ capture are
 mono-energetic electrons which have energy well separated from
 the $\beta$ decay spectrum. The background of the $\beta$ decay
 events does not affect the detection of $\nu_{R1}$. Moreover,
 this experiment does not require measurement of very high precision as
 in KATRIN {\bf ~\cite{Katrin}} or in detecting cosmic background neutrinos~\cite{cmm,lvv}.
 Rather than using gaseous Tritium in KATRIN, large volume of $^3$H or
 $^{106}$Ru target in solid state can be used in $\nu_{R1}$ capture experiment.
 This experiment might be done in the near future.
 We note that future X-ray observation may improve the constraint on
 $\theta^{'2}_1$ which will reduce the expected event rate of $\nu_{R1}$ capture.
 However the scenario of keV scale right-handed neutrino as DM candidate
 is hard to be ruled out and the DM capture by radioactive nuclei proposed in this article
 is one way to detect this keV scale DM candidate.

 We note that background events of the capture of solar pp
 neutrinos are negligible. According to the standard solar model the flux of
 pp neutrinos of energy $\lsim 10$ keV is about
 $ 8.\times 10^6$ cm$^{-2}$ s$^{-1}$~\cite{bahc} and
 the corresponding density of these pp neutrinos is about $2.7 \times
 10^{-4}$ cm$^{-3}$. The event rate of the capture of pp neutrinos
 of energy $\lsim 10$ keV is $\sim 4.0\times 10^{-3}$ year$^{-1}$
 for $10$ kg of Tritium and is $\sim 8.5 \times 10^{-2}$ year$^{-1}$ for
 $10$ ton of $^{106}$Ru.
 Effects of low energy solar neutrinos can be neglected in discussing
 the capture of $\nu_{R1}$ DM.

 Recent analysis of X-ray observations of local dwarf Willian 1 show
 evidence that the $\nu_{R1}$ DM may have mass around 5 keV with
 mixing $|R_{l1}|^2 \lsim 10^{-9}$ ~\cite{lk}. Another
 analysis of the X-ray observation of the galactic center suggests
 that $\nu_{R1}$ DM has mass around 17 keV with mixing $|R_{l1}|^2
 \sim 10^{-12}$~\cite{ps}. This small neutrino mixing would give too small
 event rates of $\nu_{R1}$ capture by radioactive nuclei unless we are living
 in a sub-halo of DM.

 {\bf Conclusion:}
 In conclusion we have considered several issues of the keV scale
 right-handed $\nu_{R1}$ DM in the $\nu$SM.
 We have shown that a GeV scale right-handed neutrino in the
 $\nu$SM can have sufficient small couplings to active neutrinos
 and its decay can lead to large amount of entropy release ( $\sim 100$ )
 at MeV temperature scale. This happens when two GeV scale right-handed
 neutrinos are degenerate. Two degenerate
 right-handed neutrinos in their interaction base can have very different
 strengths of couplings with active neutrinos and a large suppression to the
 Yukawa couplings of one of the GeV scale $\nu_R$'s can be achieved.
 Mass and mixing of light neutrinos can be correctly explained
 in the model considered. Density of $\nu_{R1}$ DM can be diluted
 by a factor larger than 100 in the reheating produced by the decay of this GeV scale
 right-handed neutrino. The mixing of $\nu_{R1}$ with active
 neutrinos is allowed to be as large as $\theta_1^2 \sim 10^{-6}$
 for $ M_1 \approx 2$ keV, reaching the bound from X-ray observation.

 We have discussed the capture of the cosmic
 background of keV scale $\nu_{R1}$ DM by radioactive nuclei
 $^3$H and $^{106}$Ru. $\nu_{R1}$ capture produces
 events of mono-energetic electrons with keV scale energy away from the beta decay spectrum.
 The background of beta decay events does not affect the detection of $\nu_{R1}$ capture events and
 this experiment does not require measurement of very high precision as
 in experiment of direct measurement of neutrino mass or in detecting cosmic background neutrinos.
 Rather than using gaseous Tritium in experiment of direct measurement of neutrino mass,
 large volume of $^3$H or
 $^{106}$Ru target in solid state can be used in $\nu_{R1}$ capture experiment.
 We find that there are about 0.7 events per year on $10$ kg tritium target
 and about 16 events per year on $10$ Ton $^{106}$Ru target
 for mixing $|R_{e1}|^2=10^{-6}$ and $n_{\nu_{R1}}=10^5$ cm$^{-3}$.

 We comment that if we are living in a sub-halo of DM the event
 rate can be much larger. Two examples with $^3$H and $^{106}$Ru
 targets are related to beta decay. It is interesting
 if suitable radioactive nucleus can be found to do $\nu_{R1}$ capture
 in anti-beta decay experiment.
 We emphasize that detecting $\nu_{R1}$ DM on radioactive nuclei is
 a new type of experiment for detecting DM in the universe.
 It is very interesting if other radioactive nucleus, better than $^3$H and $^{106}$Ru,
 can be found. This radioactive nucleus should have reasonable lifetime, large enough capture
 rate and reasonably large production rate in reactors or is available in nature.
 It's worth further exploration.
 \\

 {\bf Acknowledgement:} I would like to thank A. Yu. Smirnov for
 his encouragement. This work is supported
 by Science and Technology Commission of Shanghai Municipality under
 contract number 09PJ1403800 and
 National Science Foundation of China(NSFC), grant 10975052.


\begin{thebibliography}{99}

\bibitem{ABS} % The $\nu$MSSM, dark matter and neutrino masses
 T. Asaka, S. Blanchet and M. Shaposhnikov,
 Phys. Lett. B{\bf 631}, 151(2005).

  \bibitem{bhl}
 F. Bezrukov, H. Hettmansperger, M. Lindner, Phys. Rev. {\bf
 D81}, 085032(2010).% [arXiv:0912.4415].

 \bibitem{ci}
 J. A. Casas and A. Ibarra, Nucl. Phys. {\bf B618}, 171 (2001).
 %[arXiv:hep-ph/0103065].

\bibitem{others} A. D. Dolgov and S. H. Hansen, Astropart. Phys. {\bf 16}, 339 (2002);
K. Abazajian, G.M. Fuller, and M. Patel, Phys. Rev. {\bf D64}, 023501 (2001);
K. Abazajian, G.M. Fuller, and W.H. Tucker, Astrophys. J. {\bf 562}, 593 (2001).

\bibitem{DW} %Sterile Neutrinos as Dark Matter
 S. Dodelson and L. W. Windrow,
 Phys. Rev. Lett. {\bf 72}, 17(1994).

\bibitem{blrv} %Realistic sterile neutrino dark matter with keV mass does not
 %contradict cosmological bounds,
 A. Boyarsky, et.al., Phys. Rev. Lett. {\bf 102}, 201304(2009).

  \bibitem{bnrst} %How to find a dark matter sterile neutrino?,
 A. Boyarsky, et.al., Phys. Rev. Lett. {\bf 97}, 261302(2006).

 \bibitem{bnr} %Constraints on the parameters of radiatively decaying dark matter
 %from the dark matter halo of the milky way and ursa minor,
 A. Boyarsky, J. Nevalainen and O. Ruchayskiy, Astron. Astrophys. {\bf 471}, 51(2007).

 \bibitem{birs}
 %Constraints on decaying dark matter from XMM-Newton observations of M31,
 A. Boyarsky, D. Iakubovskyi, O. Ruchayskiy and V. Savchenko,
 MNRAS {\bf 387}, 1361(2008).

 \bibitem{blrv2}
 %Lyman-$\alpha$ constraints on warm and on warm¨Cplus¨Ccold dark matter models
  A. Boyarsky, J. Lesgourgues, O. Ruchayskiy, M. Viel,
 JCAP {\bf 0905}, 012(2009).%[arXiv:0812.0010].

 \bibitem{bri}  %A Lower bound on the mass of Dark Matter particles
 A. Boyarsky, O. Ruchayskiy, D. Iakubovskyi, JCAP {\bf 0903}, 005(2009).
 %[arXiv:0808.3902].

 \bibitem{BS} %Searching for dark matter sterile neutrinos in the laboratory
 F. Bezrukov and M. Shaposhnikov, Phys. Rev. {\bf D75}, 053005(2007).

  \bibitem{ak} % Interactions of keV sterile neutrinos with matter
 S. Ando and A. Kusenko,  Phys. Rev. {\bf D81}, 113006(2010).%[arXiv:1001.5273]

 \bibitem{sh}
 M. Shaposhnikov, Nucl. Phys. B{\bf 763}, 49(2007).%[hep-ph/0605047].

 \bibitem{hll}
 %Symmetry, dark matter, and LHC phenomenology of the minimal nuSM
 X.-G. He, T. Li and W. Liao, Phys. Rev. {\bf D81}, 033006 (2010).
 %[arXiv:0911.1598]

 \bibitem{ST} %The $\nu$MSM, inflation, and dark matter
 M. Shaposhnikov and I. Tkachev, Phys. Lett. {\bf B639}, 414 (2006).

\bibitem{Kusen} %Sterile Neutrinos, Dark Matter, and Pulsar Velocities
 % in Models with a Higgs Singlet
 A. Kusenko, Phys. Rev. Lett. {\bf 97}, 241301 (2006).

\bibitem{PK}
 %Dark-matter sterile neutrinos in models with a gauge singlet in the Higgs sector
 K. Petraki and A. Kusenko, Phys. Rev. {\bf D77}, 065014 (2008).

\bibitem{KS}
%Sterile neutrino dark matter in B . L extension of the standard model and galactic 511 keV line
 S. Khalil and O. Seto, JCAP{\bf 0810}, 024(2008).%[arXiv:0804.0336]

\bibitem{Gouvea}
%See-saw energy scale and the LSND anomaly
For other model of low energy seesaw see also A. de Gouvea, Phys.
Rev. {\bf D72}, 033005(2005). %[hep-ph/0501039].

 \bibitem{stkt}
 R. J. Scherrer and M. S. Turner, Phys. Rev. {\bf D31}, 681(1985);
 E. W. Kolb and M. S. Turner, Front. Phys. {\bf 69}, 1(1990).

 \bibitem{ask}
  T. Asaka, M. Shaposhnikov and A. Kusenko, Phys. Lett. {\bf B638},
  401(2006).

 \bibitem{brs} %The Role of Sterile Neutrinos in Cosmology and Astrophysics
 %For a recent review see
 A. Boyarsky, O. Ruchayskiy and M. Shaposhnikov, Ann. Rev. Nucl. Part. Sci. {\bf 59},
 191(2009). % [arXiv:0901.0011].

 \bibitem{cmm} %Probing low energy neutrino backgrounds with neutrino
 %capture on beta decay nuclei
 A. G. Cocco, G. Mangano and M. Messina, JCAP{\bf 0706}: 015(2007);
 J. Phys. Conf. Ser. {\bf 110}, 082014(2008).

 \bibitem{lvv} %Charged current cross section for massive
 %cosmological neutrinos impinging on radioactive nuclei
 R. Lazauskas, P. Vogel, C. Volpe, J. Phys. {\bf G35}, 025001(2008).
 %[arXiv.org: 0710.5312].

 \bibitem{bahc} Neutrino Astrophysics, J. N. Bahcall,
 Cambridge University Press 1989.

 \bibitem{Katrin}
 G. Drexlin et. al., Nucl. Phys. B(Proc. Suppl){\bf 145}, 263(2005).

 \bibitem{lk}  M. Loewenstein, A. Kusenko, Astrophys. J. {\bf 714},
 652(2010).
 %[arXiv:0912.0552].

 \bibitem{ps} D. A. Prokhorov, J. Silk, arXiv:1001.0215.

\end{thebibliography}
\end{document}